\documentclass[preprint]{elsarticle}

\usepackage{array}
\usepackage{dcolumn}
\usepackage{url}
\usepackage{breakurl}

\usepackage{bm}
\usepackage[colorlinks, pdfborder={0 0 0}, plainpages=false]{hyperref}
\usepackage{graphicx}
\usepackage{xspace}
\usepackage[usenames,dvipsnames]{color}
\usepackage{amssymb}
\usepackage[normalem]{ulem} 
\usepackage{textcomp}
\usepackage{letltxmacro}
\usepackage{amsfonts}
\usepackage{amsmath}
\usepackage{amssymb}
\usepackage{bm}
\usepackage{dcolumn}
\usepackage{epsfig}
\usepackage{graphics}
\usepackage[latin1]{inputenc}
\usepackage{latexsym}
\usepackage{rotating}
\usepackage{hyperref}
\usepackage[usenames]{color}
\usepackage{float}
\usepackage{xspace} 
\usepackage{mathrsfs}
\usepackage{subfigure}
\usepackage{multirow}
\usepackage{booktabs}
\usepackage{soul}

\usepackage{colortbl}
\usepackage{etoolbox}
\usepackage[table]{xcolor}

\def\lesssim{\mathrel{\hbox{\rlap{\hbox{\lower4pt\hbox{$\sim$}}}\hbox{$<$}}}}
\def\gtrsim{\mathrel{\hbox{\rlap{\hbox{\lower4pt\hbox{$\sim$}}}\hbox{$>$}}}}
\def\alt{\mathrel{\hbox{\rlap{\hbox{\lower4pt\hbox{$\sim$}}}\hbox{$<$}}}}
\def\agt{\mathrel{\hbox{\rlap{\hbox{\lower4pt\hbox{$\sim$}}}\hbox{$>$}}}}

\usepackage{mathtools}

\def\gta{\ifmmode {\mathbin{\lower 3pt\hbox   
    {$\,\rlap{\raise 5pt\hbox{$\char'076$}}\mathchar"7218\,$}}}
    \else {${\mathbin{\lower 3pt\hbox
    {$\rlap{\raise 5pt\hbox{$\char'076$}}\mathchar"7218\,$}}}
    $}\fi}
\def\lta{\ifmmode {\,\mathbin{\lower 3pt\hbox   
    {$\,\rlap{\raise 5pt\hbox{$\char'074$}}\mathchar"7218\,$}}}
    \else {${\mathbin{\lower 3pt\hbox
    {$\rlap{\raise 5pt\hbox{$\char'074$}}\mathchar"7218\,$}}}
    $}\fi}

\newcommand{\msun}{{\rm M}_{\odot}}
\newcommand{\beq}{\begin{equation}}
\newcommand{\eeq}{\end{equation}}
\newcommand{\bea}{\begin{eqnarray}}
\newcommand{\eea}{\end{eqnarray}}

\definecolor{darkperiwinkle}{RGB}{102, 102, 128}

\definecolor{light-gray}{gray}{0.9}

\newcommand{\RNum}[1]{\uppercase\expandafter{\romannumeral #1\relax}}

\usepackage{lineno,hyperref}
\modulolinenumbers[5]

\journal{Journal of \LaTeX\ Templates}

\begin{document}

\begin{frontmatter}

\title{Deep learning for gravitational wave forecasting of neutron star mergers}



\author[NCSA,CAII,PNCSA]{Wei Wei}
\author[NCSA,CAII,PNCSA,iCASU,ANCSA]{E. A. Huerta}

\address[NCSA]{National Center for Supercomputing Applications, University of Illinois at Urbana-Champaign, Urbana, Illinois 61801, USA}
\address[CAII]{NCSA Center for Artificial Intelligence Innovation, University of Illinois at Urbana-Champaign, Urbana, Illinois 61801, USA}
\address[PNCSA]{Department of Physics, University of Illinois at Urbana-Champaign, Urbana, Illinois 61801, USA}
\address[iCASU]{Illinois Center for Advanced Studies of the Universe, University of Illinois at Urbana-Champaign, Urbana, Illinois, 61801, USA}
\address[ANCSA]{Department of Astronomy, University of Illinois at Urbana-Champaign, Urbana, Illinois 61801, USA}


\begin{abstract}
\noindent We introduce deep learning time-series forecasting for gravitational wave detection of 
binary neutron star mergers. This method enables the identification of these signals in 
real advanced LIGO data up to 30 seconds 
before merger. When applied to GW170817, our deep learning forecasting method identifies the 
presence of this gravitational wave signal 10 seconds before merger. 
This novel approach requires a 
single GPU for inference, 
and may be used as part of an early warning 
system for time-sensitive multi-messenger searches. 
\end{abstract}


\end{frontmatter}


\section{Introduction}
\label{intro}

Multi-messenger observations of gravitational wave sources provide a wealth of information 
about their astrophysical properties and environments. For instance, gravitational wave 
and electromagnetic observations of the binary neutron star GW170817~\cite{bnsdet:2017} 
provided new insights into the equation of state of supranuclear matter, the cosmic factories 
where the heaviest r-procese elements are produced, and the progenitors of short-gamma 
ray bursts and kilonovae~\cite{mma:2017arXiv,Hurtley1551,grb:2017ApJ,kiloGW170817:2017,Mooley:2017enz,2017Natur55171T}. 
This and future multi-messenger 
observations will continue to 
advance our understanding of fundamental physics, gravitation, cosmology and 
astrophysics~\cite{Soares_Santos_2019,Abbott,Hubble_from_GW,Maya:2018F,Berti:2018GReGr,LIGOScientific:2019fpa}. 

The rationale to design and deploy a coordinated multi-messenger and multi-wavelength 
follow-up framework has been persuasively discussed in the literature~\cite{LIGOScientific:2019gag,mma_nature_revs,2019NatRP...1..585M,2019NatRP...1..600H}. The plethora of 
studies conducted for GW170817, which involved dozens of observatories in 
every continent that covered all messengers and the entire range of the electromagnetic 
spectrum, have shown that future multi-messenger discoveries depend critically on the 
development of prompt response or early warning systems to obtain a full understanding 
of astrophysical events.~\cite{Burns:2019zzo} 

To further emphasize this point, early warning systems go beyond the development 
of algorithms for real-time detection of gravitational wave sources, which could then 
be used to trigger electromagnetic and 
astro-particle follow-ups. A step beyond real-time gravitational wave detection 
consists of the development of algorithms that identify gravitational wave signals 
in real gravitational wave data before the merger takes place. Such an idea and 
implementation in the context of template-matching algorithms, using stationary Gaussian 
data recolored to advanced LIGO and advanced Virgo sensitivities, was introduced 
in~\cite{Sachdev:2020lfd}. 

In this article we introduce the use of deep learning time-series 
forecasting to identify the presence of gravitational wave signals 
in advanced LIGO data. This approach provides pre-merger alerts for 
binary neutron star mergers to facilitate prompt multi-messenger 
observation campaigns. We tested this novel approach by injecting 
modeled binary neutron star waveforms in advanced LIGO data, finding that 
deep learning forecasting is able to provide early warnings up to 30 seconds 
before merger. In the case of GW170817, deep learning forecasting provides 
an early warning 10 seconds before merger. It is worth pointing out that 
our deep learning model issues this early warning even when the data are 
contaminated by significant noise anomalies, as in the case of GW170817. 

This article is organized as follows. Section~\ref{sec:methods} describes 
the deep learning model used for time-series forecasting, the modeled 
waveforms and advanced LIGO noise used for training and testing. We 
summarize the results of this study in Section~\ref{sec:results}. We outline 
future directions of work in Section~\ref{sec:end}.


\section{Methods}
\label{sec:methods}

In this section we describe the use of time-series forecasting in the context of gravitational 
wave detection. We provide a succinct description of the waveform approximant and the 
advanced LIGO noise used to train and test these algorithms.

\subsection{Spectrograms}

\begin{figure*}[!h]
    \centerline{
    \includegraphics[width=0.5\linewidth]{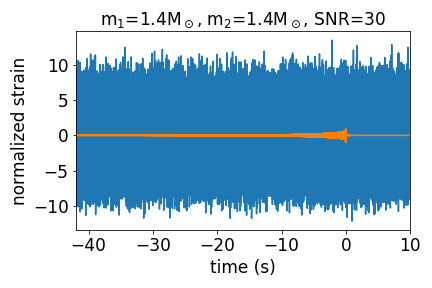}
    \includegraphics[width=0.5\linewidth]{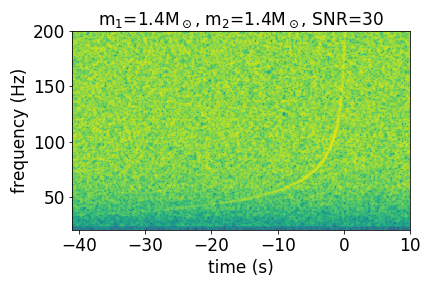}
    }
    \caption{Left panel: modeled waveform of a \((1.4\msun,\,1.4\msun)\) binary neutron star. This signal is 
    produced by the \texttt{IMRPhenomD\_NRTidal} approximant at a sample rate of 16384Hz. This modeled 
    waveform in injected in advanced LIGO's second observing run data. Right panel: spectrogram of the 
    waveform signal shown in the left panel.}
    \label{fig:spectrogram}
\end{figure*}

Spectrograms provide a visual representation of the frequencies that make up a signal 
as it evolves in time. Figure~\ref{fig:spectrogram} shows the spectrogram of a 
\((1.4\msun,\,1.4\msun)\) 
binary neutron star signal, as described by the \texttt{IMRPhenomD\_NRTidal} 
approximant~\cite{Dietrich:2018uni} 
at a sample rate of 16384Hz. This modeled waveform has been injected in advanced 
LIGO's second observing run data.

\noindent Binary neutron stars usually exhibit long-duration chirp signals, as shown 
in Figure~\ref{fig:spectrogram}. This property makes them a 
useful tool to forecast the merger event, and produce an early warning alert of an 
imminent event that may be accompanied by electromagnetic counterparts. We 
explore this idea in the following sections to produce early warnings of multi-messenger 
sources. 

\subsection{Chirp-pattern recognition with deep learning}
\label{sec:patter_recognition}

We use chirps in spectrograms to forecast the presence of gravitational wave signals in 
advanced LIGO data. We do this by training a deep neural network, 
\texttt{ResNet-50}~\cite{resnet50}, to search for chirp-like signatures in spectrograms. 
To begin with, we apply deep transfer learning to a pre-trained \texttt{ResNet-50} model, 
which is provided by \texttt{PyTorch}~\cite{NEURIPS2019_9015}. The inputs to \texttt{ResNet50} 
are the spectrograms of 8s-long advanced LIGO strain data from both the Livingston 
and Hanford observatories. These spectrograms are stacked together to form an image 
of two channels. Since the pre-trained \texttt{ResNet50} provided by \texttt{PyTorch} takes 
input images with 3 (RGB) channels, we padded the third channel of our spectrogram 
images with zeros. The output of the pre-trained \texttt{ResNet50} is a number in the 
range $[0,1]$, which indicates the probability of the presence of a chirp signal in the input spectrogram.

\subsection{Data Curation}
\label{sec:curation}

\noindent \textbf{Modeled waveforms} We use 
\texttt{PyCBC}~\cite{alex_nitz_2020_4075326} to produce 60s-long modeled 
waveforms with the \texttt{IMRPhenomD\_NRTidal} approximant at a sample rate 
of 16384Hz. We cover the parameter space \(m_{\{1,2\}}\in[1\msun,\,5\msun]\). 
The waveforms are randomly split into a training set 
(16250 waveforms) and a test set (4051 waveforms). The waveforms are 
then rescaled and injected into real advanced LIGO Livingston and Hanford noise 
to simulated binary neutron star signals over a broad range of signal-to-noise 
ratios (SNRs). 

\noindent \textbf{Advanced LIGO noise} For training we use 4 segments of open 
source advanced LIGO noise~\cite{Vallisneri:2014vxa}. These 
4096s-long segments, sampled at $16384$Hz, from the Livingston and 
Hanford observatories start with GPS times $1186725888$, $1187151872$, 
$1187569664$, and $1186897920$. The 4096s-long LIGO strain data segments 
with starting GPS time $1186725888$, $1187151872$, $1187569664$ are used 
for training, while the one with GPS starting time $1186897920$ is used for testing. 

For each of the 4096s-long LIGO strain data segments, we first calculate the 
corresponding power spectral density (PSD), and use it to whiten 
both the strain data and the waveform templates we plan to inject. We also 
rescale the amplitudes of the whitened templates and add 8s-long whitened 
LIGO strain data and templates together to simulate different SNRs. Finally, the standard 
deviation of the LIGO strained data with signal injections is normalized to one. 
The spectrograms are calculated from 8s-long simulated signals from the 
Livingston and Hanford observatories.

The spectrograms are calculated with the \texttt{spectrogram} function 
provided by \texttt{SciPy}. We use a \texttt{blackman} window size of $16384$, 
and a step size of $1024$. The resulting spectrogram is of size $8193\times 113$, 
where $8193$ is the size in the frequency domain and $113$ is the size in the time 
domain. We also apply an element-wise $\log$ transformation on the spectrograms 
to accentuate the chirp patterns.

Since the goal for the trained \texttt{ResNet-50} is to predict binary neutron star mergers 
based on information we process during the inspiral phase, we only used data from 
$20$Hz to $150$Hz on the spectrograms. This approach speeds up both the training 
and the inference. It follows that the input to \texttt{ResNet-50} is an image of 
size $130\times 113$ with three channels, where the first two channels are spectrograms 
calculated from Livingston and Hanford data, and the third channel is padded with zeros.

\subsection{Training strategy}
\label{sec:training}

The \texttt{ResNet-50} model provided by \texttt{PyTorch} is pre-trained 
with \texttt{ImageNet}~\cite{cvpr:Deng} data that spans 1000 classes. 
We changed the last fully connected layer of the default \texttt{ResNet-50} 
so that the output is a number in the range $[0,1]$, instead of an array of size 1000.

As mentioned above, we consider 60s-long signals with 
component masses \(m_{\{1,2\}}\in[1\msun,\,5\msun]\), injected in 
advanced LIGO data, and which describe a broad range of SNRs. 
Furthermore, since we 
focus on early detection, we consider the evolution of these signals as they sweep through 
a gravitational wave frequency range between \lbrack20Hz, 150Hz\rbrack. We use these datasets 
to produce a spectrogram dataset to train \texttt{ResNet-50} using a batch size of $256$, 
and a learning rate of $10^{-4}$.  

To improve the robustness of the trained \texttt{ResNet-50}, 
$50\%$ of the input spectrogram images contain no signals, while $25\%$ have 
simulated signals only in one of the Livingston and Hanford observatories, 
and the remaining $25\%$ have signals in both Livingston and Hanford strain data.
We trained the \texttt{ResNet-50} using 4 \texttt{NVIDIA V100} GPUs with 
the ADAM~\cite{kingma2014adam} optimizer.

\section{Results}
\label{sec:results}

To test the performance of our early warning model, we searched for patterns associated 
with the existence of gravitational waves in spectrograms. As mentioned above, this search cover the 
frequency range \lbrack20Hz, 150Hz\rbrack. 
We used a sliding window of 8s, with a step size of 1s, that is applied to the first 50s 
of the spectrograms. Notice that in the preparation of our training dataset, the first 50s of the 
modeled waveforms describe the pre-merger evolution. The output of our deep learning model 
provides the probability of the existence of a gravitational wave in input spectrogram.

\begin{figure*}
\centerline{
\includegraphics[width=0.5\linewidth]{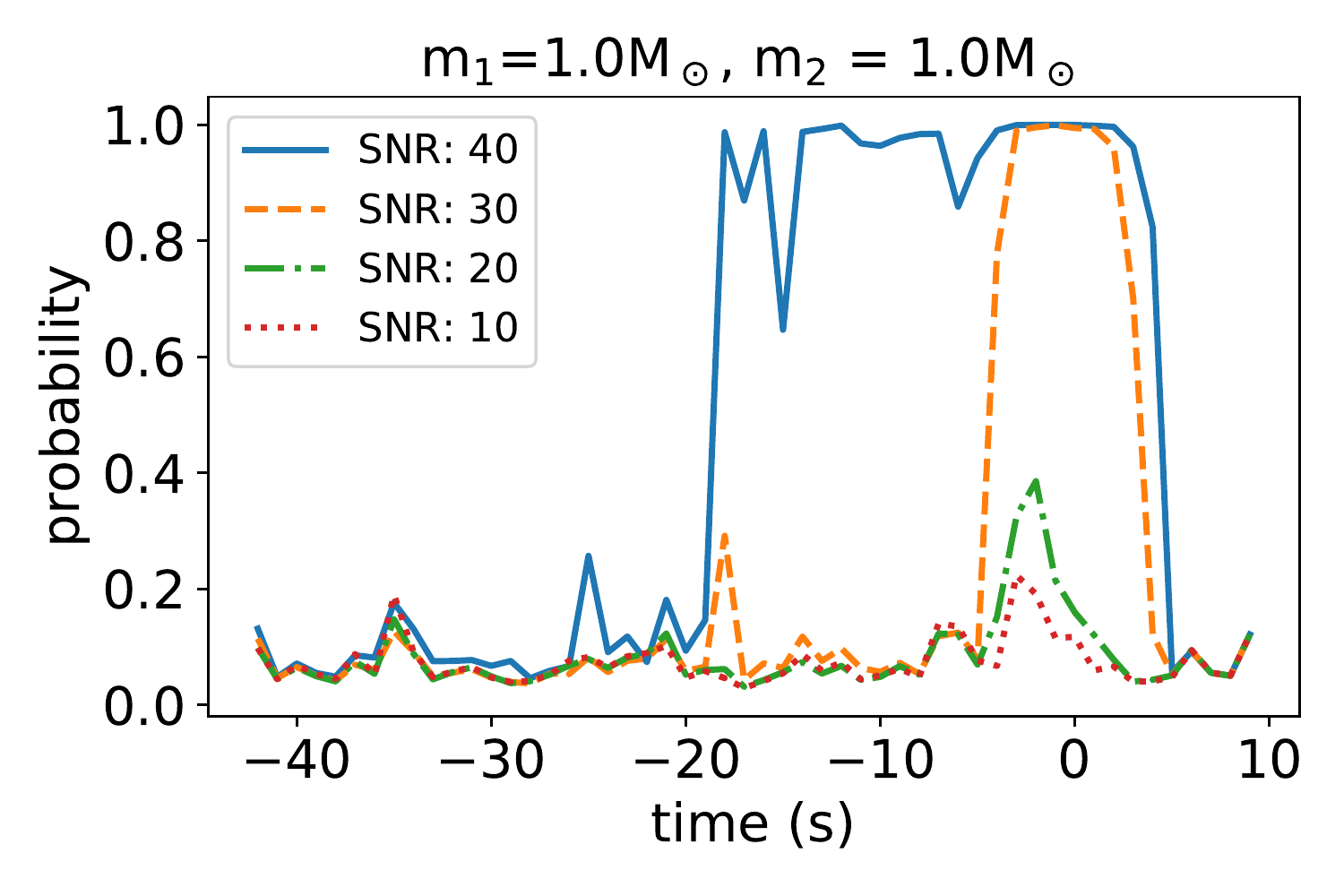}
\includegraphics[width=0.5\linewidth]{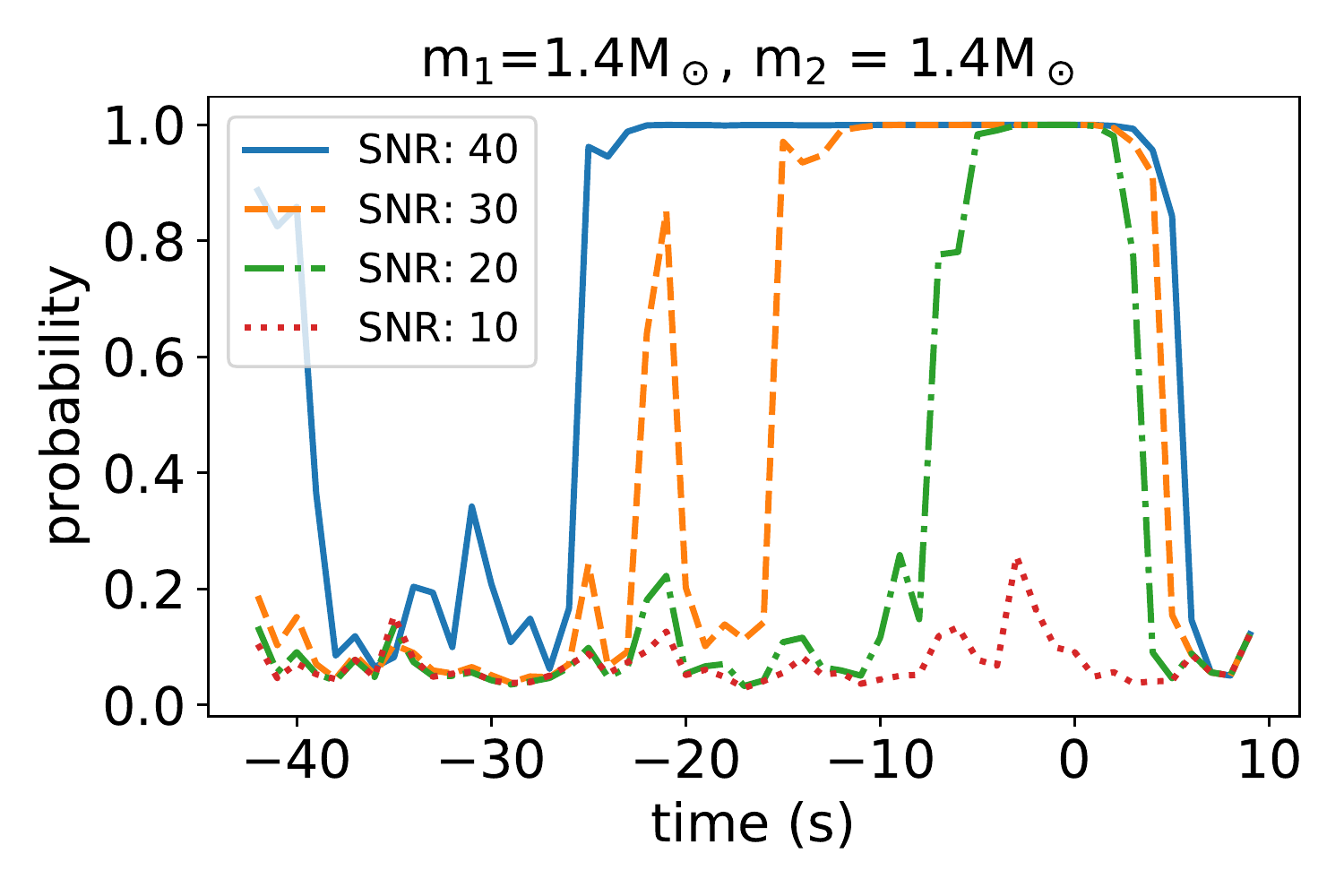}
}
\centerline{
\includegraphics[width=0.5\linewidth]{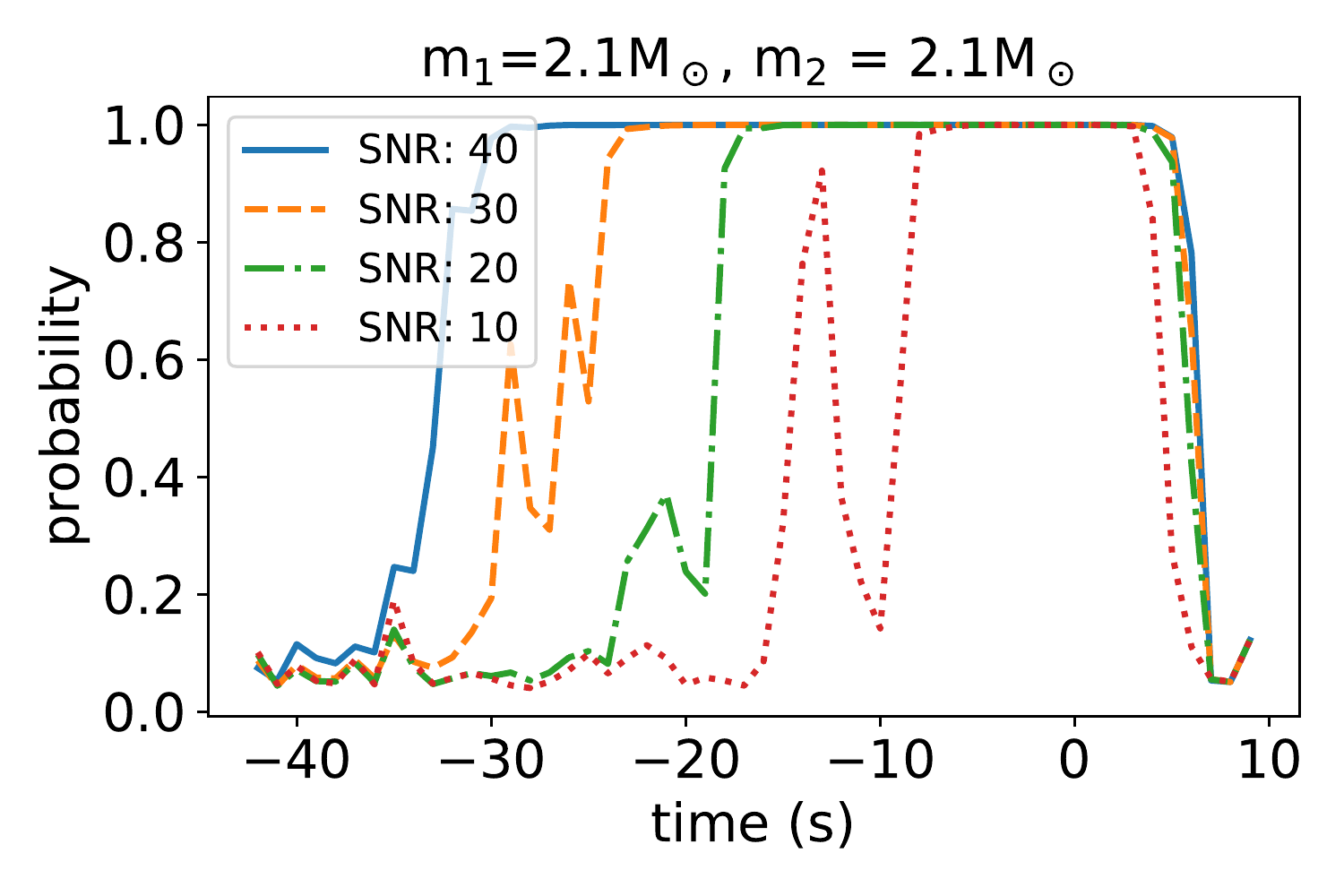}
\includegraphics[width=0.5\linewidth]{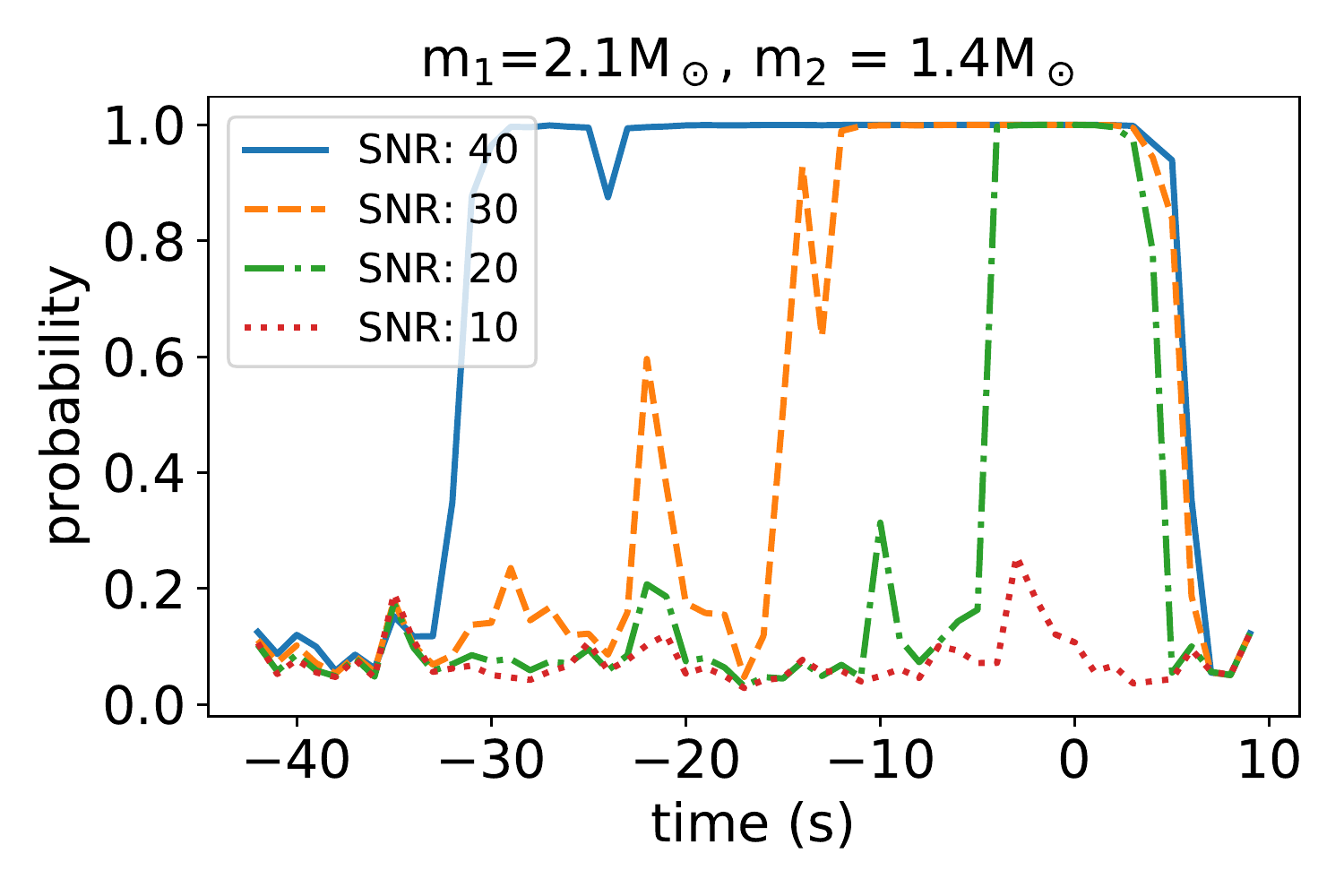}
}
\caption{Deep learning forecasting for binary neutron stars in advanced LIGO data. An astrophysically motivated 
sample of binary systems and signal-to-noise ratios show that deep learning identifies signals in real data up to 
30s before merger.}
\label{fig:bns_pre}
\end{figure*}

Figure~\ref{fig:bns_pre} presents a summary of our results. We consider four cases to 
illustrate how early our deep learning model predicts the existence of signals in advanced LIGO noise. 
We notice that deep learning forecasts the existence of binary neutron star signals in advanced 
LIGO data up to 30s before merger. As expected, the neural network performs best for signals that 
have SNRs similar to GW170817. 

\subsection{GW170817}

\begin{figure*}
    \centerline{
    \includegraphics[width=0.5\linewidth]{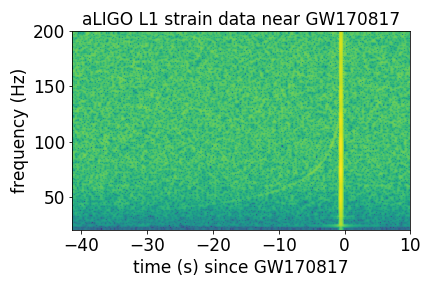}
    \includegraphics[width=0.5\linewidth]{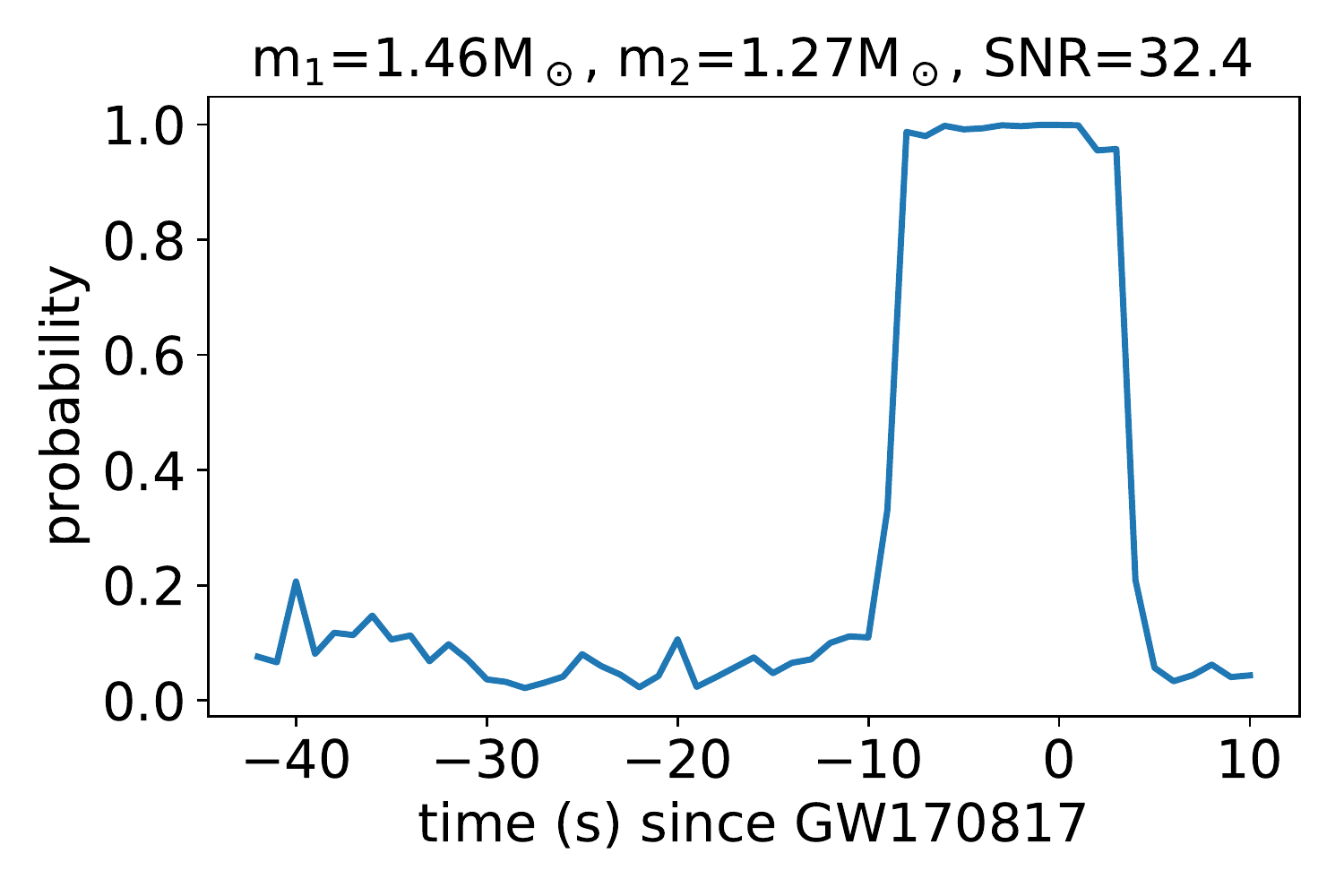}
    }
    \centerline{
    \includegraphics[width=0.5\linewidth]{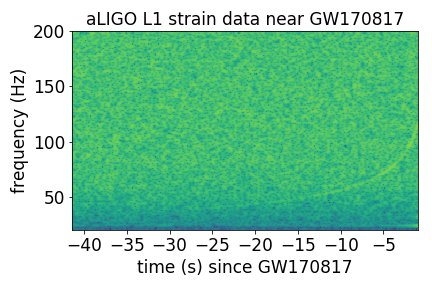}
    \includegraphics[width=0.5\linewidth]{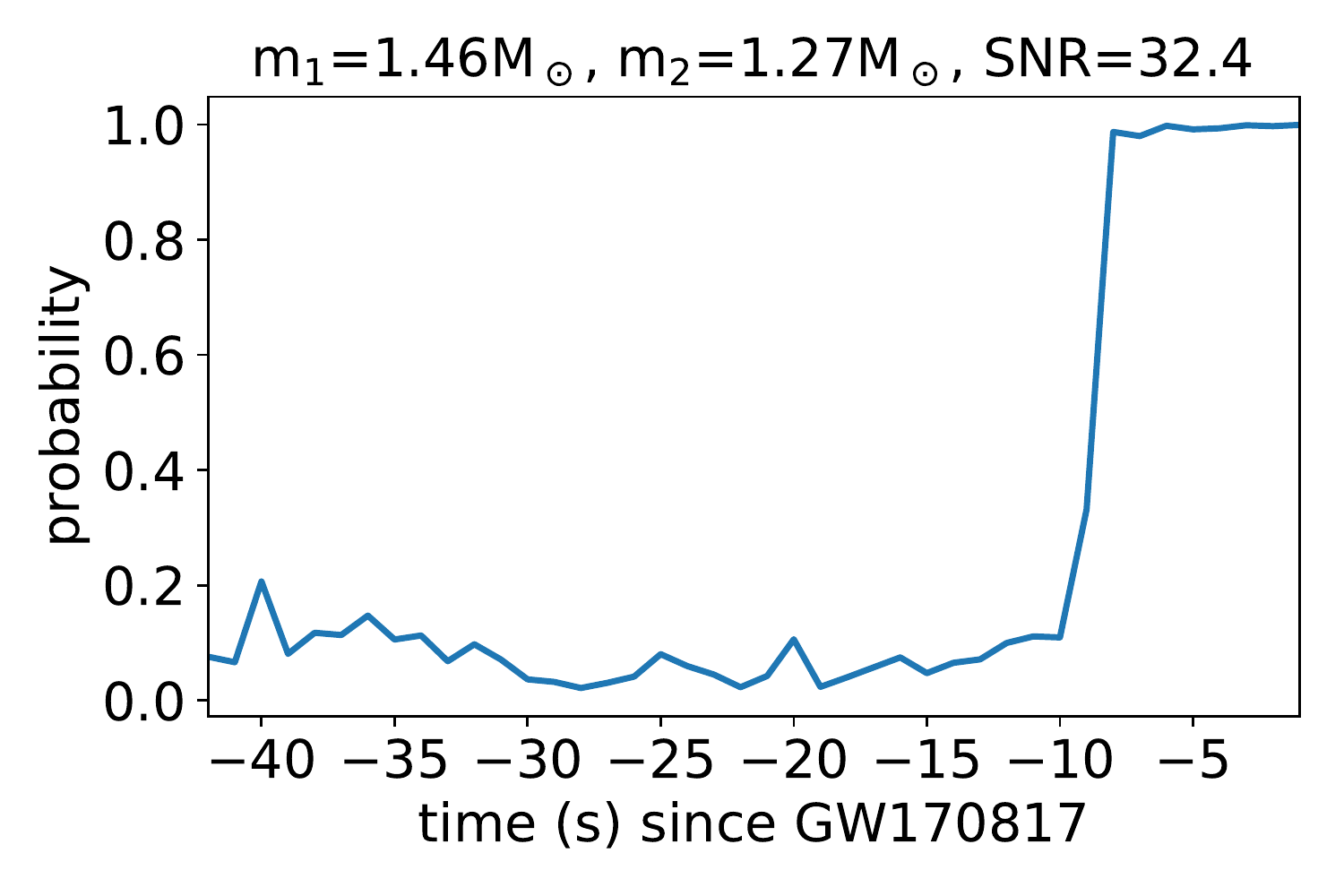}
    }
    \caption{Top panel: deep learning forecasts the existence of GW170817 ten seconds before merger. Notice that the prediction is not affected by the existence of a significant noise anomaly at merger, marked by \(t=0\mathrm{s}\) in the spectrogram. Bottom panel: our deep learning early warning system predicts the existence of GW170817 by processing real LIGO data that does not include the noise anomaly in the vicinity of \(t=0\mathrm{s}\). The top-right and bottom-right panels show that deep learning forecasting is not affected by the noise anomaly present in GW170817 data.}
    \label{fig:gw170817}
\end{figure*}

We have put at work our forecasting model in the context of GW170817 data. 
Using available, open source, advanced LIGO data for this event, we have found that 
our approach predicts the existence of this event 10s before merger. We have considered 
two datasets, one including the well known noise anomaly that contaminated this event 
(top panel in Figure~\ref{fig:gw170817}), and one without this glitch (bottom panel in 
Figure~\ref{fig:gw170817}). Our results clearly show that deep learning forecasting is not affected by 
this noise anomaly.

\subsection{Neutron star-black hole binaries}

\begin{figure*}
\centerline{
\includegraphics[width=0.5\linewidth]{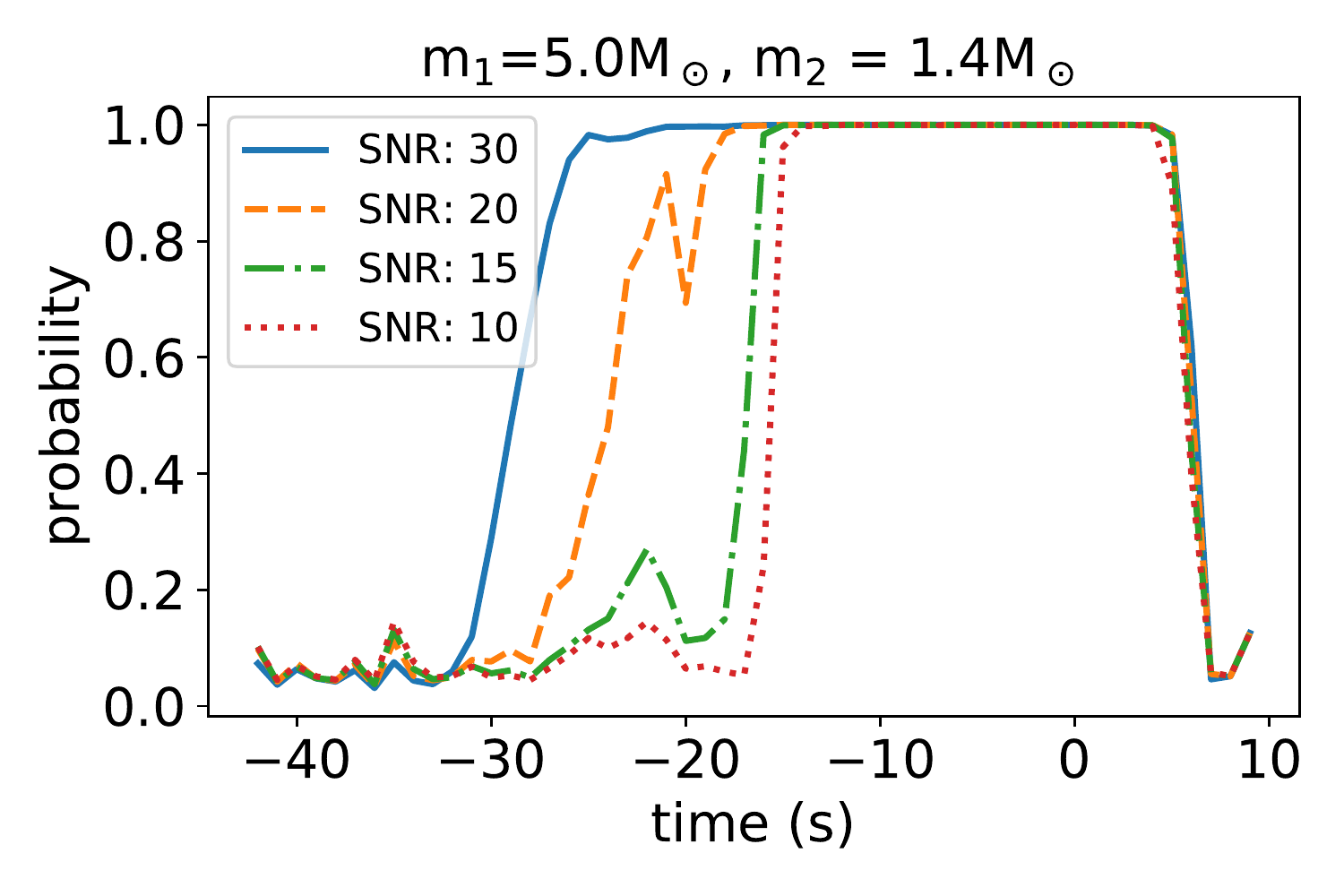}
\includegraphics[width=0.5\linewidth]{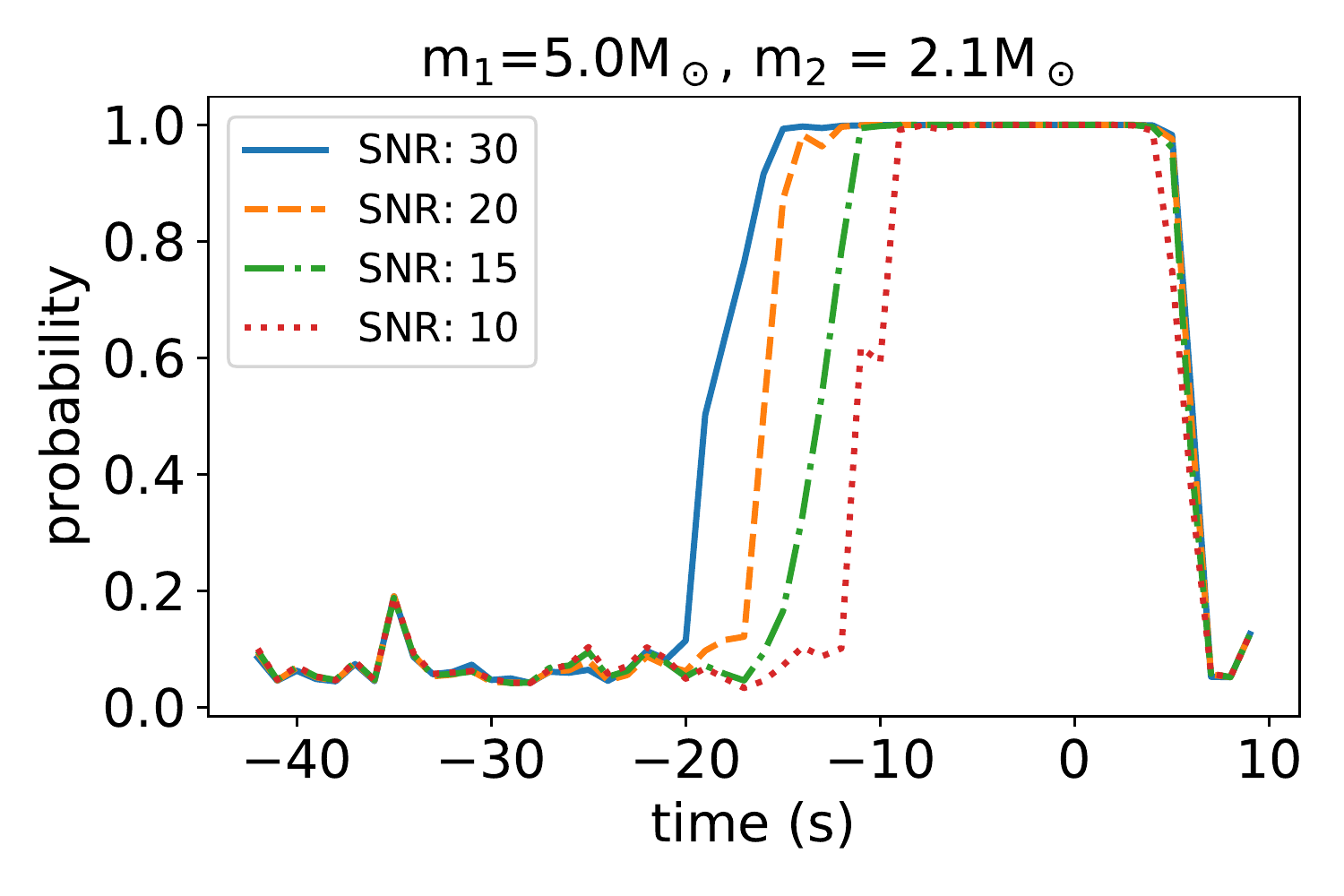}
}
\caption{Right panel: forecasting results for neutron star-black hole systems with component masses  \((5\msun,\, 1.4\msun)\)  spanning a broad range of signal-to-noise rations. Right panel: as left panel but now for \((5\msun,\, 2.1\msun)\) binaries.}
\label{fig:nsbh_pre}
\end{figure*}

We finish these analyses with an application of our early warning system in the context of 
neutron star-black hole systems. We consider two cases, that describe systems with 
component masses \((5\msun,\, 1.4\msun)\) and \((5\msun,\, 2.1\msun)\). As before, 
we inject the signals describing these binaries in advanced LIGO data, and consider a 
number of SNRs. Figure~\ref{fig:nsbh_pre} shows that our deep learning model can forecast 
the existence of these systems in advanced LIGO data up to 30s before merger for 
\((5\msun,\, 1.4\msun)\) with \(\mathrm{SNR}\sim30\), and up to 20s before merger for 
\((5\msun,\, 2.1\msun)\) with \(\mathrm{SNR}\sim30\). In other words, the method we introduce 
in this paper may be used to obtain early warnings up to 30s before the binary components 
coalesce. This information may in turn be used to enable time-sensitive electromagnetic 
follow-ups of binary neutron stars 
and neutron star-black hole systems. These results are very promising, and warrant the extension of 
this approach to other astrophysical scenarios of interest.

\section{Conclusions}
\label{sec:end}
We have introduced the first application of deep learning forecasting for the 
detection of binary neutron stars. We have also presented an application of 
this framework in the context of neutron star-black hole systems. Our results 
indicate that deep learning may provide early warnings up to 30s before merger. 

When we apply this novel methodology for GW170817, we found that 
deep learning forecasts the existence of this event 10s before merger. Our approach 
is robust to the presence of glitches, as we report for this event. 

These results lay the foundation for the construction of a deep learning forecasting method 
that can provide early warnings to enable rapid electromagnetic follow-ups. We will 
present an extended version of this framework for other astrophysical scenarios 
of interest in the near future.

\section{Acknowledgments}
\label{ack}

We gratefully acknowledge National Science Foundation (NSF) awards 
OAC-1931561 and OAC-1934757. We thank NVIDIA for their continued support. 
This work utilized resources supported by the NSF's Major Research Instrumentation 
program, the Hardware-Learning Accelerated (HAL) cluster, grant OAC-1725729, 
as well as the University of Illinois at Urbana-Champaign.


\bibliography{awareness}

\end{document}